# Sporadic Electron Jets from Cathodes – The Main Breakdown-Triggering Mechanism in Gaseous Detectors

C. Iacobaeus, M. Danielsson, P. Fonte, T. Francke, J. Ostling, V. Peskov

*Abstract*-- We have demonstrated experimentally that the main breakdown-triggering mechanism in most gaseous detectors, including micropattern gaseous detectors, is sporadic electron jets from the cathode surfaces. Depending on conditions, each jet contains randomly from a few primary electrons up to $10^5$, emitted in a time interval ranging between $0.1 \mu s$ to milliseconds. After the emission, these primary electrons experience a full gas multiplication in the detector and create spurious pulses.

The rate of these jets increases with applied voltage and very sharply at voltages close to the breakdown limit. We found that these jets are in our measurements responsible for the breakdown-triggering at any counting rate between $10^{-2}$ $Hz/mm^2$ and $10^8$ $Hz/mm^2$.

We demonstrated on a few detectors that an optimized cathode-geometry, a high electrode surface quality and a proper choice of the gas mixture, considerably improve the performance characteristics and provide the highest possible gains.

## I. INTRODUCTION

IN previous work [1] we have studied some basic properties of various micropattern gaseous detectors. For example the maximum achievable gain in single and multi-step configurations was studied, and also the gain vs. rate characteristics. We have in the present work focused the study on primary reasons of breakdowns: what is actually the main mechanism triggering the breakdown in gaseous detectors, especially when there is no external radiation?

## II. EXPERIMENTAL SETUP

Our main experimental setup is shown in Fig. 1. It contains a test chamber inside of which various gaseous detectors can be installed.

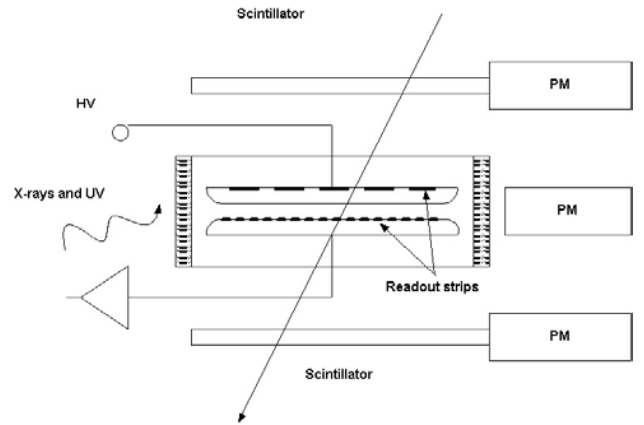

Fig. 1. A schematic drawing of the experimental setup.

We have for these studies chosen the most "popular"/most often used, gaseous detectors: The Single Wire Counter (SWC), the Multi Wire Proportional Counter (MWPC), the Micro Strip Gaseous Counter (MSGC), the Parallel Plate Avalanche Chamber (PPAC), the Resistive Plate Chamber (RPC), the Micromesh Gaseous Detector (MICROMEGAS) and the Gas Electron Multiplier (GEM). Schematic drawings of some of these designs are presented in Fig. 2(a), 2(b), 2(c) and 2(d).

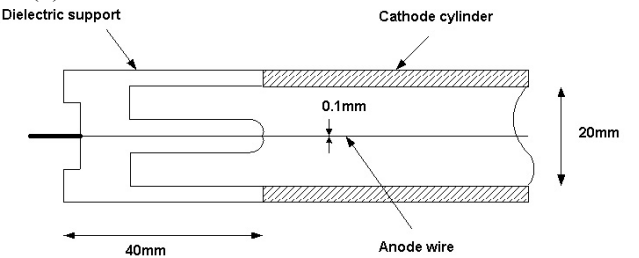

Fig. 2(a). A schematic drawing of the SWC.

These were carefully designed to minimize, or to possibly fully avoid, contribution of spurious pulses and breakdowns associated with the dielectrics supports in the anode and the cathode interfaces. The design of the interface in the case of the SWC, the MWPC, the PPAC and the RPC (made of

C. Iacobaeus is with the Dept. of Medical Radiation Physics, Karolinska Institutet, Box 260, 171 76 Stockholm, Sweden (telephone: +46-8-5537 8188, e-mail: christian@radfys.ks.se).

M. Danielsson is with the Dep. of Physics, Royal Institute of Technology, SCFAB, KTH Particle Physics, 106 91 Stockholm, Sweden (telephone: +46-8-5537 8181, e-mail: mats.danielsson@mamea.se).

P. Fonte is with the ISEC and LIP, Coimbra University, Coimbra P-3000, Portugal (e-mail: paulo.fonte@cern.ch).

T. Francke is with the Dep. of Physics, Royal Institute of Technology, SCFAB, KTH Particle Physics, 106 91 Stockholm, Sweden (telephone: +46-8-5537 8180, e-mail: tom.francke@xcounter.se).

J. Ostling is with the Dept. of Medical Radiation Physics, Stockholm University, Box 260, 171 76 Stockholm, Sweden (telephone: +46-8-5537 8188, e-mail: janina@radfys.ks.se).

V. Peskov is with the Dep. of Physics, Royal Institute of Technology, SCFAB, KTH Particle Physics, 106 91 Stockholm, Sweden (telephone: +46-8-5537 8182, e-mail: vladimir.peskov@cern.ch).

Pestov-glass of $10^{10}\,\Omega cm$ or Si-plates of $2\cdot 10^{3}\,\Omega cm$) can be seen in Fig. 2(a), 2(b) and 2(c).

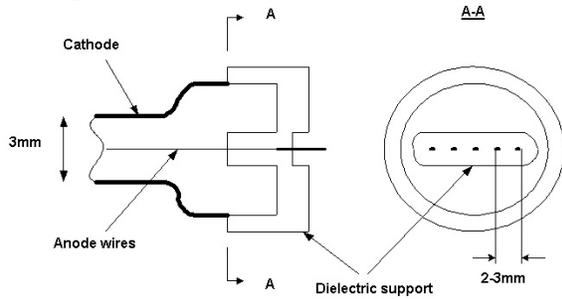

Fig. 2(b). A schematic drawing of the MWPC.

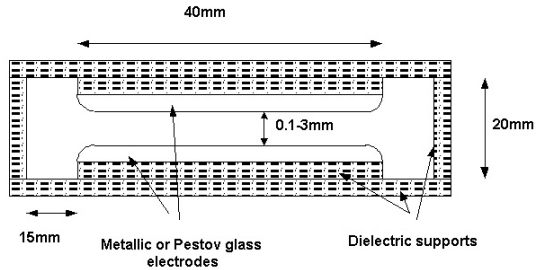

Fig. 2(c). A schematic drawing of the PPAC/RPC.

In the case of the MICROMEGAS, we used an anode plate with metallic readout strips of 50 $\mu m$-pitch. This setup allows the readout strips to be placed far away from the spacers (Fig. 2(d)), thus only recording avalanches in this active area and avoiding the signals due to possible micro-breakdowns across the spacer's surface.

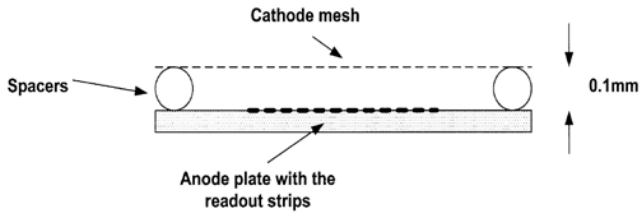

Fig. 2(d). A schematic drawing of the MICROMEGAS.

The MSGC used in this experiment had a size of 10 cm x 10 cm and was made of Desag-glass with Cr-strips of a 200 $\mu m$-width. They were manufactured at the Photomask Inc. in USA.

The GEMs (10 cm x 10 cm) were obtained from CERN and had a standard design with a hole-diameter of 70 $\mu m$ and a pitch of 140 $\mu m$.

We have additionally studied large areas (25 cm x 25 cm and 25 cm x 100 cm) glass-RPC of $10^{12}\,\Omega cm$, described in [2].

Before installation in the test chambers, all detectors were cleaned very carefully by ultrasonic bath in various solvents, acetone and alcohol. The final cleaning was also done ultrasonically in a soap and distilled water solution, and then in distilled water only. The tests were performed in different Ar-, Xe- and Kr-based mixtures with various combinations of Isobutane, Freon (R134), Freon + Isobutane, Ethane and $CO_2$ at atmospheric pressure. The ratio of each component was varied widely.

In studies at low counting rates ($< 10^4\,Hz/mm^2$), the ionization inside the detectors was caused by x-rays (x-ray gun of 6-30 keV), betas ($^{90}Sr$ and $^{106}Ru$) and gammas ($^{60}Co$). UV-light from a mercury lamp was used in order to create single electrons from the cathodes. The efficiency of the detectors for minimum ionizing particles was measured using cosmic muons. They were identified by coincidence signals from two scintillators (see Fig. 1).

For position measurements, we used a G10 readout plate with 20 metallic strips of 1 cm-pitch attached to the outer surface of the RPC. In an avalanche mode of operation, signals from the RPCs were measured at low rates with charge sensitive amplifiers, and at high rates with current amplifiers. In streamer mode, signals were directly monitored on the LeCroy oscilloscope. In the case of measuring position resolutions, signals from the strips were simultaneously monitored by seven two-channeled storage LeCroy oscilloscopes.

We additionally used PM-tubes facing the windows of the test chambers in some measurements in order to detect the light produced by avalanches and streamers inside the detectors (see Fig. 1).

In contrast to our previous work [1], the measurements in high counting rates ($> 10^5\,Hz/mm^2$) were done with modulated sources. In the case of x–rays, the modulation was done with a mechanical shutter being able to fully close the beam at reaction time of ~0.01 s. We also used powerful pulsed gamma radiation produced by the Racetrack accelerator at the Karolinska Hospital (see Fig. 3 and [3]).

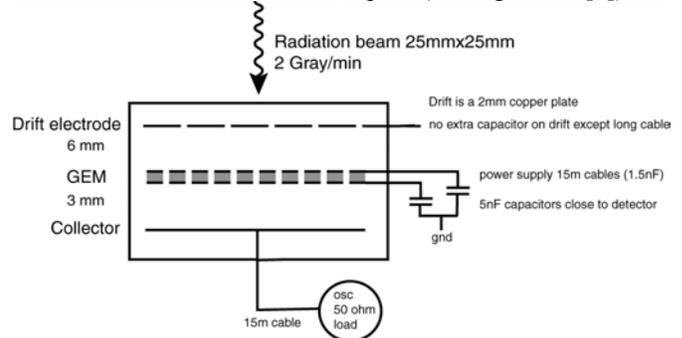

Fig. 3 A schematic drawing of the test chamber with a GEM inside for measurements at the Racetrack.

The Racetrack (MM50) produces a scanned electron beam (50 MeV) hitting a target consisting of mainly Tungsten. Hence a beam of photons with a maximum energy of 50 MeV (average of 20 MeV) is produced, and the purpose of the accelerator is radiation treatment of cancer patients. The radiation beam is pulsed ($\sim 5\mu s$ long pulses) with about 5 ms pauses between pulses. The photon fluence rate during the tests was $2\,Gy/min$ or correspondingly about $10^{10}\,photons/mm^2 \cdot s$ during the pulse.

## III. RESULTS

### A. Weak external radiation ($< 10^4$ Hz/mm$^2$) and no external radiation

As an example, Fig. 4 and 5 show typical dependence of rate vs. voltage for pulses produced with and without an external radiation source for the MWPC and the RPC. The number of radiation-induced pulses typically increases with voltage and then reaches a plateau. In contrast to this, the rate of the spurious pulses always increases rapidly with the applied voltage. This behavior is typical for all detectors tested in this work. The question we try to address in this work regards the origin of these pulses.

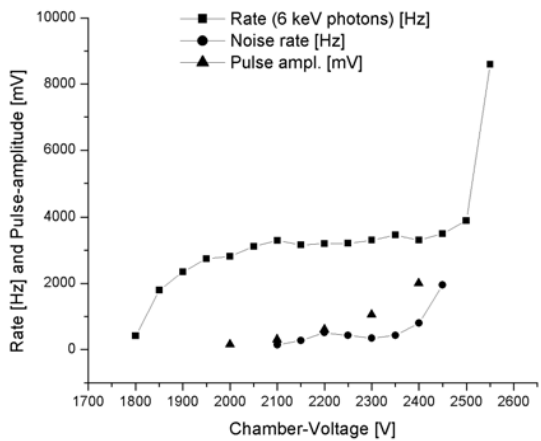

Fig. 4. Signal amplitude and rates of noise pulses and pulses produced by x-rays as a function of the voltage measured with the MWPC.

Since the rate of the spurious pulses, and sometimes their amplitudes, are much higher than the rate and amplitudes of pulses due to cosmic radiation or natural radioactivity, they are the main triggers of breakdown at low intensity of the external radiation. This was proven experimentally by measuring spurious pulses and breakdown rates in coincidence measurements of cosmic muons. As an example, Fig. 5 shows the rate of spurious pulses, and these were measured using the coincidence technique.

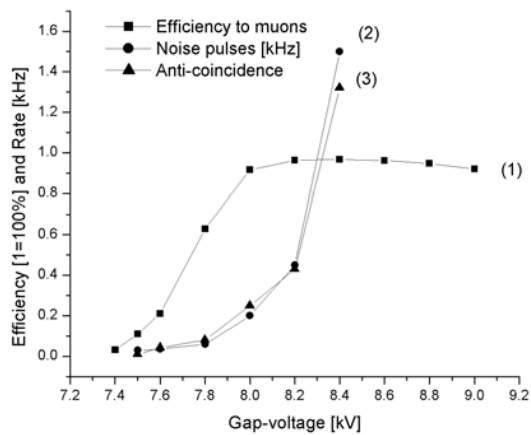

Fig. 5. The efficiency (1) and the rate of noise pulses (2), (3) vs. the voltage applied on the RPC [2]. (2) and (3) correspond to measurements done in coincidence and anti-coincidence with the signals from the scintillators respectively.

More detailed studies reveal that most of the detectors (except in the case of the GEM), have only two types of spurious pulses: one distributed randomly with time (at relatively low voltages) and the other clustered in time (usually at voltages close to breakdown).

Some light was shed upon this problem by experiments with PPACs and RPCs. As an example Fig. 6(a), 6(b), 6(c) and 6(d) show an appearance of clustered spurious pulses from the RPC. The upper trace of the oscillogram shows the pulse from the PM-tube coupled to the scintillator (triggered on muons in coincidence with the other scintillator). The lower trace shows a pulse due to a muon (directly measured on the 50 ohm-input of the oscilloscope) from the RPC, and also afterpulses appearing with an increasing applied voltage. The amplitudes of these afterpulses are very randomly distributed and may be considerably larger than the muon pulses. Note that many other authors have observed these afterpulses [4].

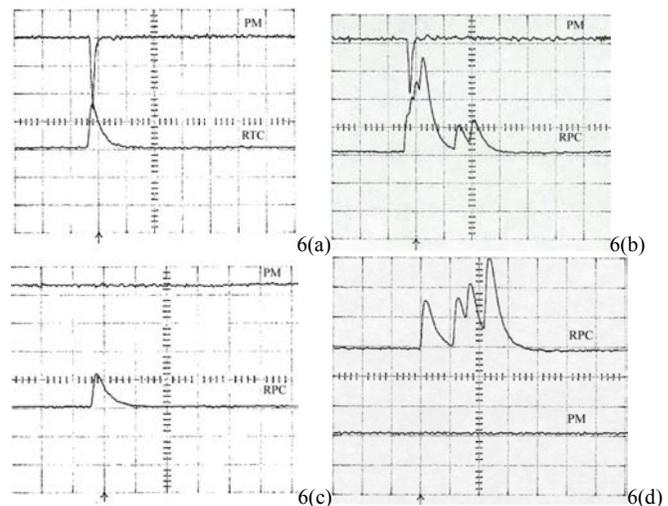

Fig. 6. Signals measured in coincidence with cosmic muons (a), (b) and the noise pulse (c), (d) at various voltages V applied on the RPC [2], (a) and (c) at V=7.6 kV and (b) and (d) at V=8,75 kV. The gas mixture used was Ar/Isobutane/Freon (R134) in the ratio 48/4/48.

The common explanation to the origin of the afterpulses is that they are due to a photo-effect caused by a primary avalanche or a streamer inside the detector [4]. However, our observations show that grouped spurious pulses have a sporadic delay that sometimes can be very long, and this excludes the explanation based on the photo-effect. By comparison of the pulse height spectra of the spurious pulses with those produced by single photo-electrons (measured with detectors operating in proportional mode), one can conclude that the spurious pulses, depending on conditions, may contain between a few up to a few thousands of electrons [5].

### B. The role of aging and other types of depositions

We have found that any depositions on metallic cathodes, e.g. polymer films due to aging or dust particles, cause an increase in the rate of noise pulses. These observations were independently confirmed by other authors, see for example [6].

*1) The "Memory" effect*

We have discovered during these studies that all tested detectors exhibit what we call a "memory" effect. The memory effect manifests itself in two ways:

1) The value of the safe operating voltage (no sparking) should be decreased with an increasing counting rate.

2) After a breakdown, one has to reduce the voltage on the detector some value, $\Delta V$, for some period of time, $\Delta t$, to avoid continuos sparking.

The first effect is illustrated in Fig. 7, in which the pulse amplitude vs. voltage for MICROMEGAS is presented. One can see that at very low rates, the amplitude may correspond to as much as $10^7$ electrons per avalanche before a breakdown appears. However, even at such relatively low rates as $10$ $Hz/mm^2$, the maximum achievable gain drops. Note that each avalanche acts completely independently from each other at this low rate, since the ion's removal time from the amplification gap is about 0.1 $\mu s$. Thus the detector somehow "remembers" the action of an avalanche for a time that is an order of magnitude longer than the ion removal time.

The second effect was the need to considerably lower the voltage to avoid continuos sparking after a breakdown. The voltage could be restored only after some period of time, which varied depending on the detector and the gas mixture, usually from fractions of a second up to a few hours. In rare cases (especially in mixtures with Isobutane) one had to wait for a day.

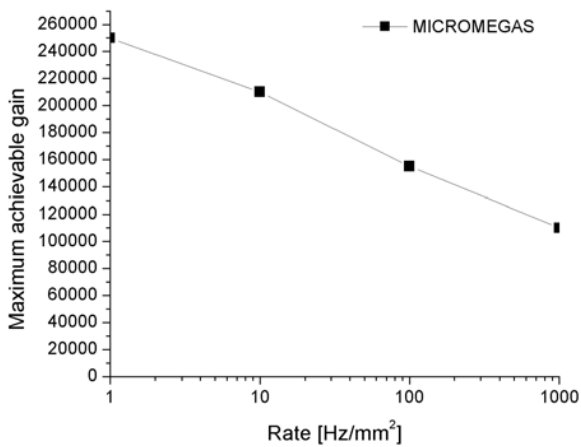

Fig. 7. The maximum achievable gain (6 keV x-rays) vs the rate for the MICROMEGAS.

*C. Results with modulated x-ray and gamma radiation*

One possible explanation of the effects described above (noise/spurious pulses and the memory effect) is that they are due to the Malter-type explosive emission from microscopic dielectric insertions or thin dielectric layers on the cathode surface (see below for more details). If there is a flux of positive ions to the cathode, these insertions or layers will be charged up and create an extremely high electric field, enough to cause a field emission effect in form of sporadic jets of electrons, [7] and [8]. One of the most straightforward ways to verify this hypotheses is to "switch off" the positive ion flux very fast. In this case, the insertions and dielectric layers will remain charged for some time and one can thus expect a continuation of electron emission for this period of time. We performed measurements with modulated x-rays and pulsed gamma radiation sources to verify this.

*1) Results with the modulated x-rays source: afterpulses*

Fig. 8 shows the rate of spurious pulses after the beam was blocked for two detectors, the PPAC and the RPC. One can see that their pulse rate decay time may reach between ten and a few hundred seconds respectively. Similar results, but with different time scales were obtained for all tested detectors. The shortest decay time (a few seconds) was observed with a well-cleaned MSGC.

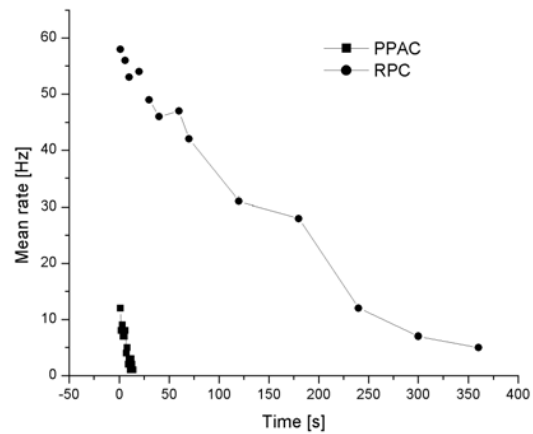

Fig. 8. The rate of the afterpulses for the PPAC (Cu-electrodes) and the RPC (Si).

*2) Results with pulsed gamma radiation*

The response of the GEM, for low voltages applied on its electrodes (350 V), to a high flux of gamma radiation from the Racetrack accelerator is presented in Fig. 9. The GEM was in these measurements loaded directly on the oscilloscope with a 50 ohm-input, without any other restrictive-current resistors.

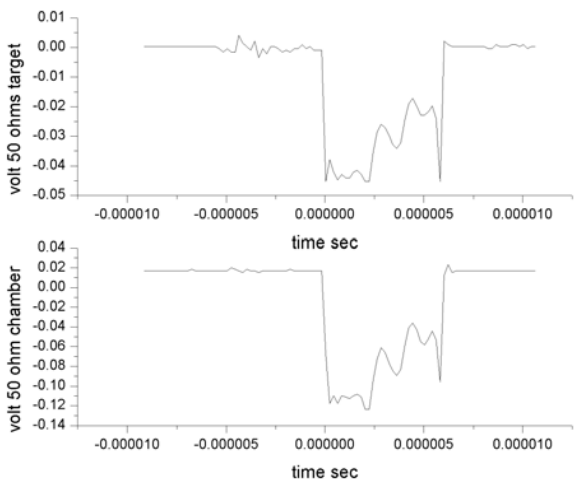

Fig. 9. The current from the GEM (at 350 V) recorded directly on a 50 ohm-input of the oscilloscope when the GEM was exposed to a pulsed gamma radiation, producing $\sim 10^7$ $counts/mm^2$ on the whole GEM-area.

The response of the GEM changed dramatically when higher voltages were applied. Fig. 10 shows a pre-breakdown-phenomena in the GEM at 420 V. Without current-restrictive resistors, a real breakdown is fatal and would fully destroy the GEM. The GEM-pulses in Fig. 10, however, are not real discharges, but jets of electrons after they multiplied in the GEM. At higher voltages, these jets cause real breakdowns. Another interesting effect that was observed was the current from the GEM increasing with time (see Fig. 10) similar to what was observed in [9]. Such a current increase is very typical for the Malter-effect [10].

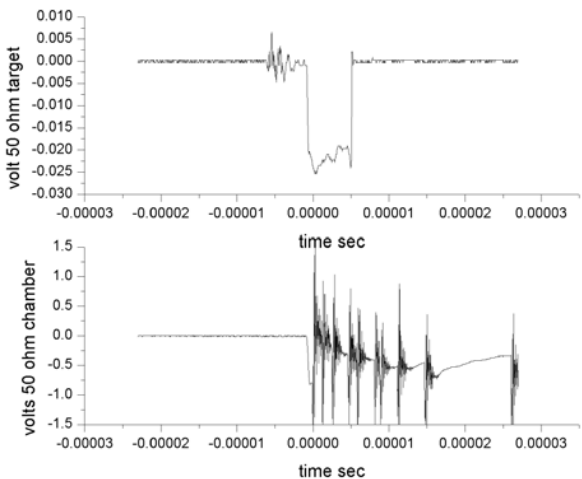

Fig. 10. The same setup as in Fig. 9, but 420 V applied over the GEM electrodes.

IV. DISCUSSION

A generally accepted explanation of the Malter-effect is a charging up of dielectric films on the metallic cathode surface by positive ions. If the dielectric film is thin enough, the created electric field may be sufficient to cause a field emission. A classical field emission predicts an emission in form of single electrons.

Studies of breakdown mechanisms in high vacuum [7] reveal however that the field emission could rather be in a form of busted electron emission, so called explosive field emission [7]. This emission originates at some points on the cathode where there are sharp tips or, even more important, microscopic dielectric insertions.

The theory of this effect is based on the fact that dielectric insertions are not ideal dielectrics and contain a system of low-energy levels. In a high electric field, electrons from the cathode are able of tunneling to the dielectric insertion where they accumulate. After some critical concentration, they suddenly emit to the vacuum in form jets of electrons.

From previous work [8] together with the recent measurements, it looks like a similar phenomenon may occur in gaseous detectors. Since the number of primary electrons emitted by the jets occasionally can be $> 10^5$ these jets will satisfy the Raether limit [8], $A \cdot n_0 > 10^8$ electrons, even at rather low gas gains ($< 10^3$). Thus at low intensity of external radiation (at low counting rates) these jets may dominate in the breakdown-triggering mechanism.

It is known that at high counting rates, the maximum achievable gain dramatically drops in the presence of heavy ionizing particles [11]. From this point of view, if jets occasionally emit $> 10^5$ electrons, they are as significant as heavy ionizing particles and could therefore be the main breakdown-triggering mechanism also at high rates. Jets may dominate breakdown at any rate, thus, in the case of extremely clean electrodes, cosmic and radioactive contamination will also contribute to breakdown, but surface contamination gives in real life the major contribution. This is because it is quite impossible to have surfaces without these insertions [7].

V. OPTIMIZATIONS

In real experiments, especially at very high rates typical for tracking measurements or medical imaging, it is quite impossible to fully avoid sparking. They can originate from heavy ionizing particles or by jets described above. Some experiments now accept detectors (for example MICROMEGAS [11]) able to withstand some sparks, but it is in this case very important that the detector recover very fast after a spark, otherwise the data acquisition could be seriously disturbed.

In terms of the definitions introduced above, the fast recovery corresponds to small values of $\Delta t$ and $\Delta V$ (see Section III.B.1). These values could be reduced using extremely clean surfaces (no oxide layers or dust particles) and gases without adsorbed layers, as was demonstrated earlier [8]. It is possible to achieve a short recovery time ($\Delta t$ in the range of seconds) in $CO_2$-based mixtures.

We also demonstrated in this work that by optimizing the cathode geometry and the anode-cathode interfaces (see Fig. 2(a), 2(b) and 2(c)), having a high electrode surface quality and choosing a proper gas-mixture ($CO_2$-based), it is possible to reach the limit of the gas gain determined by a general gain vs. rate curve (see [8] for more details). At the same time, the values of $\Delta t$ and $\Delta V$ could be considerably reduced. For example, the values obtained for the RPC, made of the commercially available Si and being well cleaned, were 10 s and 800 V respectively, when the RPC operated in a Xe+20% $CO_2$ gas-mixture at atmospheric pressure. However, after etching the Si-surface (removal of a thin oxide layer), the values became $\Delta t = 1s$, $\Delta V = 200V$ and this dramatically improved the detector performance at rates of $10^5$ $Hz/mm^2$.

## VI. Conclusion

We demonstrated on a few detectors that an optimized cathode-geometry, a high electrode surface quality and a proper choice of the gas mixture, allows one considerably to improve the performance characteristics and reach the highest possible gains. Therefore, results of these studies have a great practical importance.